\documentclass[aps,prl,twocolumn,groupedaddress,floatfix]{revtex4}

\usepackage{graphicx}
\usepackage{dcolumn}
\usepackage{bm}
\usepackage{amsmath}

\begin{document}

\title{Three-body recombination at large scattering lengths in an ultracold atomic gas}

\author{Tino Weber}
\author{Jens Herbig}
\author{Michael Mark}
\author{Hanns-Christoph N\"agerl}
\author{Rudolf Grimm}

\affiliation{Institut f\"ur Experimentalphysik, Universit\"at
Innsbruck, Technikerstr.\ 25, A-6020 Innsbruck, Austria}

\date{\today}

\begin{abstract}

We study three-body recombination in an optically trapped
ultracold gas of cesium atoms with precise magnetic control of the
s-wave scattering length $a$. At large positive values of $a$, we
measure the dependence of the rate coefficient on $a$ and confirm
the theoretically predicted scaling proportional to $a^4$.
Evidence of recombination heating indicates the formation of very
weakly bound molecules in the last bound energy level.

\end{abstract}


\maketitle

Three-body recombination, the process of two particles forming a
compound while interacting with a third particle, is of general
importance in many-body systems. Understanding three-body
processes has for a long time been a challenge in many areas of
physics. In dilute ultracold atomic gases, where unique
experimental access to interparticle interactions is available,
three-body recombination leads to the formation of diatomic
molecules. The possibility to trap and cool such molecules and
thus to realize ultracold molecular ensembles \cite{Masnou2001a}
or atom-molecule mixtures \cite{Donley2002a} holds great prospects
for future research in the field of quantum gases.

The main experimental challenge for measuring three-body
recombination is to distinguish three-body losses from two-body
losses. In magnetic traps, this is in general difficult, as
dipolar relaxation is present as an inherent two-body loss
mechanism. In the special case of magnetically trapped $^{87}$Rb,
experiments could nevertheless provide reliable data because of
the anomalously weak two-body decay in this species
\cite{Burt1997a,Soding1999a}. As a more general approach, optical
traps allow to store atoms in the lowest internal state where
two-body loss is fully suppressed for energetic reasons
\cite{Stamper-Kurn1998b}. Three-body recombination in bosonic
systems with large $s$-wave scattering length $a$ has been
explored in experiments utilizing Feshbach resonances in $^{23}$Na
\cite{Stenger1999c} and $^{85}$Rb \cite{Roberts2000a}. The results
demonstrate the dramatic enhancement of loss processes near such
resonances.

In the case of large positive $a$, theoretical studies
\cite{Fedichev1996b,Nielsen1999a,Esry1999b,Bedaque2000a} predict
formation of molecules in a weakly bound $s$ level and deduce
scaling of the three-body recombination rate with $a$ to the
fourth power. Recombination is discussed in terms of the event
rate $\nu_{\text{rec}}=\alpha_{\text{rec}}n^3$ per unit volume and
time, where $n$ denotes the particle density. In the low-energy
limit, the universal scaling law for the characteristic parameter
$\alpha_{\text{rec}}$ is given as $\alpha_{\text{rec}} = C\hbar
a^4/m$ \cite{Fedichev1996b}, where $m$ is the mass of the atom,
and $C$ a dimensionless factor. Predictions for $C$ give values
between 0 and $\sim$70
\cite{Fedichev1996b,Nielsen1999a,Esry1999b,Bedaque2000a}, with
oscillatory behavior between $C=0$ and $C_{\text{max}}=67.9$
expected for strong variations of $a$ on a Feshbach resonance
\cite{Bedaque2000a}. Previously available experimental data on
three-body recombination for various species roughly agree with
the general trend of a universal $a^4$ scaling, with evidence for
deviations from this scaling law near a Feshbach resonance
\cite{Stenger1999c}.

In this work, we employ an optical trap to measure three-body
recombination in an ultracold thermal gas of $^{133}$Cs in the
$6^2$S$_{1/2}, F=3, m_F=3$ absolute ground state. In this state,
the scattering length varies strongly with an applied magnetic
field $B$ through a combination of broad and narrow Feshbach
resonances at low field strengths \cite{Cheng2000,Leo2000}. The
values of $a$ as a function of $B$ have been calculated to high
precision \cite{Julienne:2002a}. This tunability allows us to
conduct experiments in a range of precisely controlled large
positive values of $a$, where the condition of recombination into
a weakly bound $s$ level is well fulfilled. Detailed measurements
of the time evolution of both atom number and temperature allow us
to quantify both three-body loss and heating.

The three-body loss rate coefficient $L_3$ is related to
$\alpha_{\text{rec}}$ via $L_3 = n_l \alpha_{\text{rec}}$, where
$n_l$ denotes the number of atoms lost from the trap per
recombination event. Therefore, the central relationship for
comparing measurements to theory is
\begin{equation}
\label{l3eqn} L_3 = n_l C \frac{\hbar}{m}a^4.
\end{equation}

In the recombination process the molecular binding energy
$\varepsilon$ is set free as kinetic energy. The molecule and the
third atom receive $\varepsilon/3$ and $2\varepsilon/3$,
respectively. Since usually $\varepsilon$ is large compared to the
trapping potential depth, both are expelled from the trap, setting
$n_l=3$. However, the binding energy of the weakly bound last
energy level of the dimer is given by $\varepsilon =
{\hbar^2}/{\left(m a^2\right)}$ \cite{LandauLifshitz}, and at very
large scattering length values $\varepsilon$ may be below the trap
depth and the third atom cannot escape. If the potential of the
atom trap does not confine the molecule, the dimer is lost and
$n_l=2$. If, however, the molecule is trapped and stays within the
atom cloud, it may quickly quench its high vibrational excitation
in an inelastic collision with a fourth trapped atom. The large
amount of energy released in this situation expels the collision
partners, making $n_l=3$. In either case, the kinetic energy of
the remaining atom is distributed in the ensemble, giving rise to
recombination heating.

Our measurements are conducted in a nearly isotropic optical
dipole trap formed by two crossed 100~W CO$_2$-laser beams focused
to a waist of 600~$\mu$m \cite{Weber2003a}. A vertical magnetic
field gradient of 31~G/cm cancels gravity for the $6^2$S$_{1/2},
F=3, m_F=3$ absolute ground state of Cs. The optical trapping
forces are so weak that no other $m_F$ substate can be held
against the gravitational force, thus perfect spin polarization of
the sample is enforced by the trap itself. We have measured the
trap frequencies by exciting small oscillations along the
principal axes, the geometric average results to
$\bar{\omega}/(2\pi)=14.5\pm 0.5$~Hz. The trap depth is $k_B\times
12$~$\mu$K; $k_B$ denotes the Boltzmann constant. Trap heating is
extremely weak, from long-time measurements we can give an upper
bound for the heating rate of $0.2$~nK/s. The background gas
limited $1/e$ lifetime is $\sim$$180$~s.

After loading the dipole trap \cite{Weber2003a}, the ensemble
cools down by plain evaporation within the first 10~s to
$3\times10^6$~atoms at $T=1$~$\mu$K, $1/12$ of the trap depth. To
avoid further evaporative loss and cooling in our measurements, we
cool the sample well below this limit. This is achieved by forced
radio frequency evaporative cooling in the magnetic levitation
field. We apply a radio frequency resonant with the Zeeman
splitting between two $m_F$ substates. In the vertical levitation
gradient this acts as height selective removal from the trap,
effecting 1D evaporative cooling. Evaporation proceeds at a bias
field of $B=75$~G to tune the scattering length to 1200~$a_0$
($a_0=0.053$~nm denotes Bohr's radius), which at the low initial
peak density of $3\times10^{11}$~cm$^{-3}$ keeps the elastic
scattering rate at $\sim$$200$~s$^{-1}$, high enough for efficient
cooling. A simple 7~s two-segment linear sweep of the radio
frequency provides $1.3\times 10^6$ atoms at 450~nK (peak
phase-space density $D\approx4\times10^{-3}$). Some measurements
have been taken at 200~nK starting temperature; to reach this, we
appended another linear 5~s rf ramp, leaving $4\times 10^5$ atoms
in the trap ($D\approx10^{-2}$). Temperature and atom number are
measured by taking an absorption image with typically 50~ms of
expansion in the levitation field after switching off the CO$_2$
lasers. For the expansion, the bias field is in all measurements
set to 75~G to ensure constant conditions.

\begin{figure}
\includegraphics{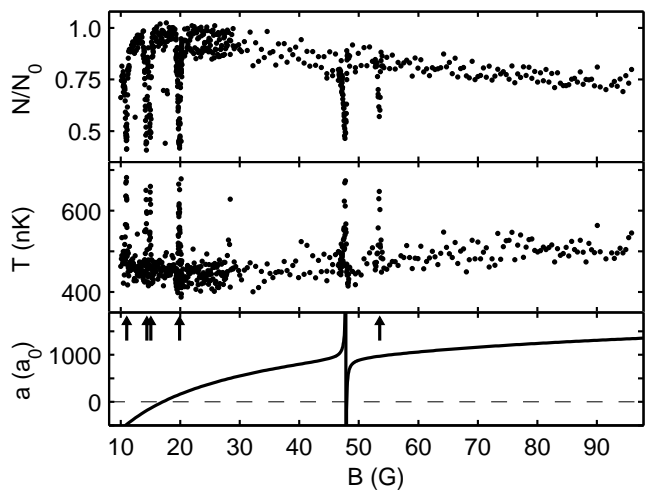}
\caption{\label{biasscan}
Remaining fraction of atoms $N/N_0$ (upper curve) and temperature
$T$ (middle) after 10~s in the CO$_2$-laser trap in a range of
magnetic bias fields. Initially $N_0=1.3\times10^6$ atoms are
prepared at $450$~nK. For comparison, the lower plot shows the
calculated scattering length $a$ with additional Feshbach
resonances indicated by arrows (see text). }
\end{figure}

In a first set of measurements, we have recorded the atom number
and temperature of the trapped sample after a holding time of 10~s
at fixed magnetic fields varied between 10~G and 100~G
{\cite{footnote_below10G}\nocite{Weber2003a}}.
Figure~\ref{biasscan} shows the data in comparison to the
calculated scattering length \cite{Julienne:2002a}, which includes
Feshbach resonances involving $s$- and $d$-wave channels.
Additional resonances involving closed $g$-wave channels are
marked by arrows. The Feshbach resonances have been observed in
previous experiments through light-induced two-body losses
\cite{Cheng2000,Cheng2001} and described theoretically
\cite{Leo2000,Cheng2001}. In our measurements, all resonances
appear as prominent loss and heating features. Over the full
range, both atom number and temperature show a clear correlation
with the magnitude of the scattering length.

\begin{figure}
\includegraphics{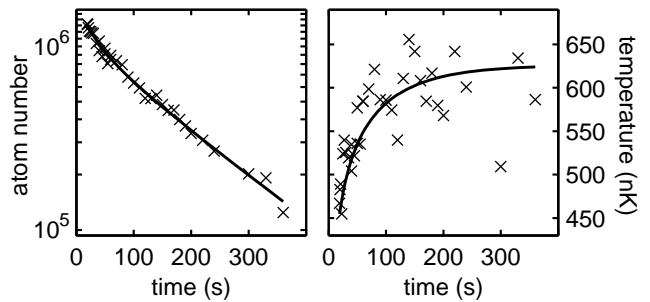}
\caption{\label{decaycurve}
Atom number and temperature as a function of time in the trap at a
bias field of 56~G. The solid lines show a numerical fit to the
data. The corresponding three-body loss coefficient is
$L_3=7(2)\times10^{-25}$~cm$^6/$s. }
\end{figure}

To obtain more quantitative results on recombination loss and
heating, we have studied the time evolution of the atom number $N$
and temperature $T$ at various magnetic fields.
Figure~\ref{decaycurve} shows a typical measurement at $B=56$~G.
Integrating the local three-body loss rate $L_3 n^3$ over the
sample gives $\dot{N}/N = -L_3\left< n^2\right>$. Expressing the
average density in terms of the directly accessible quantities $N$
and $T$, we replace $L_3\left< n^2\right>$ by $\gamma{N^2}/{T^3}$,
where $\gamma = L_3\bigl({m\bar{\omega}^2}/{2\pi
k_B}\bigr)$$^3/{\sqrt{27}}$. With an additional loss term $-\alpha
N$ for background gas collisions, we obtain a differential
equation for the atom number,
\begin{equation}
\label{diffeqN} \frac{dN}{dt} = -\alpha N - \gamma
\frac{N^3}{T^3}.
\end{equation}

Atom loss through three-body recombination leads to
``anti-evaporation'' heating. With its $n^3$ dependence,
three-body recombination predominantly happens in the region of
highest density in the trap center. The mean potential energy
\cite{footnote_kineticenergy} of an atom undergoing a three-body
process can be calculated by integrating over a thermal
distribution with weights proportional to $n^3$, which yields
$\frac{1}{2}k_B T$. As the ensemble average is $\frac{3}{2}k_B T$,
for each lost atom an excess energy of $1 k_B T$ remains in the
sample. To account for recombination heating, we introduce an
additional constant energy $k_B T_h$ per lost atom. Relating the
total heating energy $k_B(T+T_h)$ to the average energy $3k_BT$ of
a trapped particle yields
$\dot{T}/T=(\dot{N}/N)$~$k_B(T+T_h)/(3k_B T)$, and we obtain an
expression for the temperature evolution,
\begin{equation}
\label{diffeqT} \frac{dT}{dt} = \gamma \frac{N^2}{T^3} \frac{\left(T +
T_h\right)}{3}.
\end{equation}

Equations (\ref{diffeqN}) and (\ref{diffeqT}) form a set of two
coupled nonlinear differential equations for $N$ and $T$. Using an
iterative approach {\cite{footnote_iteration}\nocite{Matlab}}, we
do a least-squares fit of the numerical solutions of both
equations to the experimental data. For the data in
Fig.~\ref{decaycurve}, the fit is plotted as solid lines. Fit
parameters are $\alpha$, $\gamma$ and $T_h$. Error estimates for
the fit parameters are based on varying one parameter up and down,
resp., to the points where $\chi^2$ increases by one, while
optimizing the other parameters \cite{Bevington}.

\begin{figure}
\includegraphics{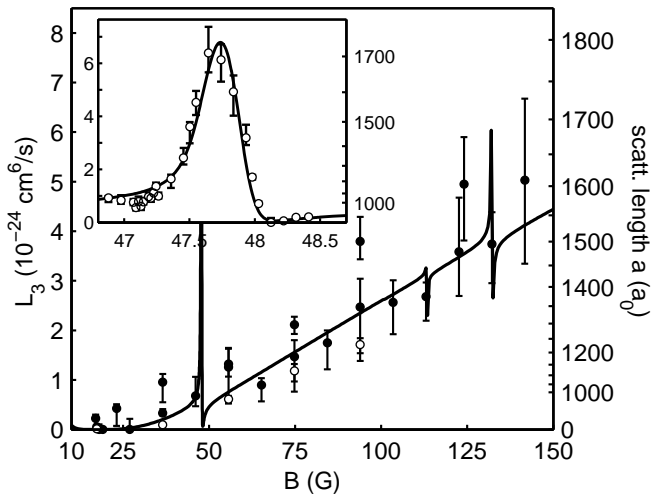}
\caption{\label{ahoch4}
Measured values of $L_3$ as a function of the bias field $B$.
Filled symbols represent measurements taken at an initial
temperature of 200~nK, open symbols at 450~nK. The solid line
shows the calculation according to Eq.~(\ref{l3eqn}) with $n_l C =
225$. The scale on the right hand side displays the corresponding
scattering length values. The inset shows the region around the
48~G resonance with a model fit (see text). }
\end{figure}

The data obtained for $L_3$ as a function of magnetic field is
plotted in Fig.~\ref{ahoch4}. The inset shows an expanded view of
data taken on the 48~G Feshbach resonance. The error bars
represent the statistical errors derived from the fit as described
above, which typically range around 20\%. The systematic
uncertainty is larger: Due to the scaling of $L_3$ with
$N^2\bar\omega^6$, small errors in these values amount to large
differences in the result obtained for $L_3$. Taking the
uncertainty in the trap frequency and an estimated 30\% in atom
number, the systematic error can be up to a factor of two.

In order to compare our data to theory, we do a least-squares fit
of the expected scaling law [Eq.~(\ref{l3eqn})] to our
measurements, using the calculated $a(B)$ \cite{Julienne:2002a}.
Single fit parameter is $n_l C$. The resulting curve is drawn as a
solid line in Fig.~\ref{ahoch4}. The data clearly confirm the
universal $a^4$ scaling. In our experimental range, where $a$
varies by less than a factor of two, we do not find any indication
of resonant behavior in $C$
{\cite{footnote_cosc}\nocite{Bedaque2000a}}. The result from the
fit is $n_l C = 225$, corresponding to $C=75$ for $n_l=3$ or
$C=112$ for $n_l=2$. In comparison to theory, both values are
somewhat above the upper limit of $C_{\text{max}}\approx 70$, but
in good agreement within our systematic error limit.

The inset in Fig.~\ref{ahoch4} shows the data on the 48~G Feshbach
resonance. In order to model the shape of the loss feature, we
have to take into account the width of the sample in the vertical
$B$ field gradient and the unitarity limitation at the finite
temperature. The theory curve is convoluted with a gaussian of
full $1/e$ width 260~mG, which is given by the vertical extension
of the $n^3$ distribution of the trapped sample. A cutoff in $a$
accounts for the unitarity limitation; the cutoff value of
1800~$a_0$ results from a least-squares fit to the data. Our
$B$-field scale is slightly adjusted for an optimum match with
theory by introducing a 40~mG shift, which is well within our
calibration accuracy.

Via the condition $ka\ll1$, the inverse wavenumber
$k^{-1}=\bigl[{{4 k_B T m}/({\pi\hbar^2})}\bigr]$$^{-1/2}$
characterizes the crossover to the zero-energy limit. At 450~nK,
$k^{-1}=1500$~$a_0$; at 200~nK, $k^{-1}=2250$~$a_0$.
Correspondingly, the 450~nK data on the Feshbach resonance show a
strong effect from the unitarity limitation, while the data taken
at 200~nK starting conditions follow the expected scaling law up
to the maximum $a\approx1500$~$a_0$.

\begin{figure}
\includegraphics{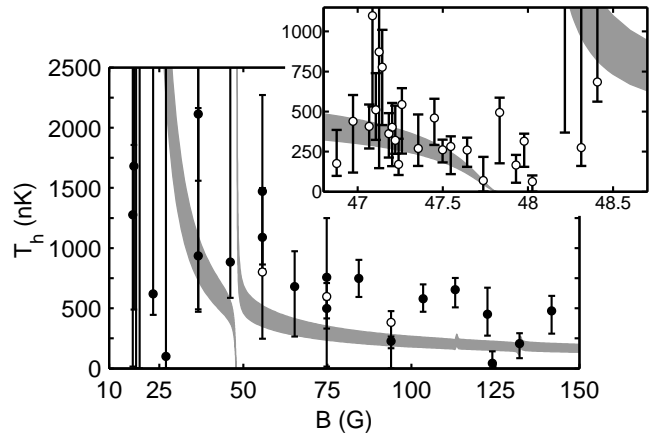}
\caption{\label{Theat}
Recombination heat $T_h$ as a function of the bias field $B$.
Filled symbols represent measurements taken at an initial
temperature of 200~nK, open symbols at 450~nK. The inset shows the
region around the 48~G resonance. The shaded area shows the
expected range based on the calculated binding energy of the last
bound state of the Cs$_2$ molecule (see text). }
\end{figure}

Our results on recombination heating are displayed in
Fig.~\ref{Theat}. A trend towards smaller $T_h$ at higher magnetic
fields, corresponding to higher scattering lengths $a$, is
visible. At large values of $a$ close to the Feshbach resonance,
the values are correspondingly very small. We compare the data to
the scenario that the recombination heating arises just from the
binding energy $\varepsilon$ of the last bound state of the dimer.
As the atom involved in the three-body recombination process
receives $2\varepsilon/3$, the recombination heat per lost atom
results to $k_B T_h =2\varepsilon/(3n_l)$. Therefore, we assume
$k_B T_h$ to be between $2\varepsilon/9$ ($n_l=3$) and
$\varepsilon/3$ ($n_l=2$). This range is shown as shaded area in
Fig.~\ref{Theat}~{\cite{footnote_epscorr}\nocite{Gribakin1993a}}.
This simple model shows reasonable agreement with the experimental
data. Our heating measurements thus provide evidence of the
formation of very weakly bound ultracold molecules.

From the apparent effect of recombination heating, we can draw an
important conclusion for the efficiency of evaporative cooling
towards quantum degeneracy in gases with large positive scattering
length. If temperatures are very low ($k_BT\ll\varepsilon$) and
the atom that carries $2\varepsilon/3$ cannot immediately escape
from the trap, recombination heating has a detrimental effect much
worse than mere three-body loss. For efficient evaporation it is
thus imperative to avoid the hydrodynamic collision regime and to
implement a three-dimensional evaporation scheme. In our cooling
experiments on optically trapped cesium \cite{Weber2003a} these
conditions indeed turned out to be essential for the attainment of
Bose-Einstein condensation.

The molecules formed in recombination are likely to remain trapped
in our setup. A previous experiment shows evidence of trapping of
the Cs$_2$ dimer in a CO$_2$-laser trap \cite{Takekoshi:1998}.
Away from Feshbach resonances, the magnetic dipole moment of the
dimer in its last bound state has the same value as the sum of the
dipole moments of two single Cs atoms, so that the levitation trap
can hold the molecules under the same conditions as the atoms.
This offers an intriguing way to study atom-molecule interactions
both in ultracold thermal clouds and Bose-Einstein condensates.
Moreover, the Feshbach resonances allow to change the molecular
magnetic moment in a controlled way and thus to separate the
molecules from the atoms to prepare pure molecular samples.

\begin{acknowledgments}
We thank C. Greene for useful discussions and P. Julienne for
providing us with the scattering length data. We gratefully
acknowledge support by the Austrian Science Fund (FWF) within SFB
15 (project part 16) and by the European Union in the frame of the
Cold Molecules TMR Network under contract No. HPRN-CT-2002-00290.
\end{acknowledgments}


\begin{thebibliography}{25}
\expandafter\ifx\csname natexlab\endcsname\relax\def\natexlab#1{#1}\fi
\expandafter\ifx\csname bibnamefont\endcsname\relax
  \def\bibnamefont#1{#1}\fi
\expandafter\ifx\csname bibfnamefont\endcsname\relax
  \def\bibfnamefont#1{#1}\fi
\expandafter\ifx\csname citenamefont\endcsname\relax
  \def\citenamefont#1{#1}\fi
\expandafter\ifx\csname url\endcsname\relax
  \def\url#1{\texttt{#1}}\fi
\expandafter\ifx\csname urlprefix\endcsname\relax\def\urlprefix{URL }\fi
\providecommand{\bibinfo}[2]{#2}
\providecommand{\eprint}[2][]{\url{#2}}

\bibitem[{\citenamefont{Masnou-Seeuws and Pillet}(2001)}]{Masnou2001a}
\bibinfo{author}{\bibfnamefont{F.}~\bibnamefont{Masnou-Seeuws}}
  \bibnamefont{and} \bibinfo{author}{\bibfnamefont{P.}~\bibnamefont{Pillet}},
  \bibinfo{journal}{Adv. At. Mol. Opt. Phys.} \textbf{\bibinfo{volume}{47}},
  \bibinfo{pages}{53} (\bibinfo{year}{2001}).

\bibitem[{\citenamefont{Donley et~al.}(2002)\citenamefont{Donley, Claussen,
  Thompson, and Wieman}}]{Donley2002a}
\bibinfo{author}{\bibfnamefont{E.~A.} \bibnamefont{Donley}},
  \bibinfo{author}{\bibfnamefont{N.~R.} \bibnamefont{Claussen}},
  \bibinfo{author}{\bibfnamefont{S.~T.} \bibnamefont{Thompson}},
  \bibnamefont{and} \bibinfo{author}{\bibfnamefont{C.~E.}
  \bibnamefont{Wieman}}, \bibinfo{journal}{Nature}
  \textbf{\bibinfo{volume}{417}}, \bibinfo{pages}{529} (\bibinfo{year}{2002}).

\bibitem[{\citenamefont{S{\"{o}}ding et~al.}(1999)\citenamefont{S{\"{o}}ding,
  Gu{\'{e}}ry-Odelin, Desbiolles, Chevy, Inamori, and Dalibard}}]{Soding1999a}
\bibinfo{author}{\bibfnamefont{J.}~\bibnamefont{S{\"{o}}ding}},
  \bibinfo{author}{\bibfnamefont{D.}~\bibnamefont{Gu{\'{e}}ry-Odelin}},
  \bibinfo{author}{\bibfnamefont{P.}~\bibnamefont{Desbiolles}},
  \bibinfo{author}{\bibfnamefont{F.}~\bibnamefont{Chevy}},
  \bibinfo{author}{\bibfnamefont{H.}~\bibnamefont{Inamori}}, \bibnamefont{and}
  \bibinfo{author}{\bibfnamefont{J.}~\bibnamefont{Dalibard}},
  \bibinfo{journal}{Appl. Phys. B} \textbf{\bibinfo{volume}{69}},
  \bibinfo{pages}{257} (\bibinfo{year}{1999}).

\bibitem[{\citenamefont{Burt et~al.}(1997)\citenamefont{Burt, Ghrist, Myatt,
  Holland, Cornell, and Wieman}}]{Burt1997a}
\bibinfo{author}{\bibfnamefont{E.} \bibnamefont{Burt}},
  \bibinfo{author}{\bibfnamefont{R.} \bibnamefont{Ghrist}},
  \bibinfo{author}{\bibfnamefont{C.} \bibnamefont{Myatt}},
  \bibinfo{author}{\bibfnamefont{M.} \bibnamefont{Holland}},
  \bibinfo{author}{\bibfnamefont{E.} \bibnamefont{Cornell}},
  \bibnamefont{and} \bibinfo{author}{\bibfnamefont{C.}
  \bibnamefont{Wieman}}, \bibinfo{journal}{Phys. Rev. Lett.}
  \textbf{\bibinfo{volume}{79}}, \bibinfo{pages}{337} (\bibinfo{year}{1997}).

\bibitem[{\citenamefont{Stamper-Kurn et~al.}(1998)\citenamefont{Stamper-Kurn,
  Andrews, Chikkatur, Inouye, Miesner, Stenger, and
  Ketterle}}]{Stamper-Kurn1998b}
\bibinfo{author}{\bibfnamefont{D.~M.} \bibnamefont{Stamper-Kurn}},
  \bibinfo{author}{\bibfnamefont{M.~R.} \bibnamefont{Andrews}},
  \bibinfo{author}{\bibfnamefont{A.~P.} \bibnamefont{Chikkatur}},
  \bibinfo{author}{\bibfnamefont{S.}~\bibnamefont{Inouye}},
  \bibinfo{author}{\bibfnamefont{H.-J.} \bibnamefont{Miesner}},
  \bibinfo{author}{\bibfnamefont{J.}~\bibnamefont{Stenger}}, \bibnamefont{and}
  \bibinfo{author}{\bibfnamefont{W.}~\bibnamefont{Ketterle}},
  \bibinfo{journal}{Phys. Rev. Lett.} \textbf{\bibinfo{volume}{80}},
  \bibinfo{pages}{2027} (\bibinfo{year}{1998}).

\bibitem[{\citenamefont{Stenger et~al.}(1999)\citenamefont{Stenger, Inouye,
  Andrews, Miesner, Stamper-Kurn, and Ketterle}}]{Stenger1999c}
\bibinfo{author}{\bibfnamefont{J.}~\bibnamefont{Stenger}},
  \bibinfo{author}{\bibfnamefont{S.}~\bibnamefont{Inouye}},
  \bibinfo{author}{\bibfnamefont{M.~R.} \bibnamefont{Andrews}},
  \bibinfo{author}{\bibfnamefont{H.-J.} \bibnamefont{Miesner}},
  \bibinfo{author}{\bibfnamefont{D.~M.} \bibnamefont{Stamper-Kurn}},
  \bibnamefont{and} \bibinfo{author}{\bibfnamefont{W.}~\bibnamefont{Ketterle}},
  \bibinfo{journal}{Phys. Rev. Lett.} \textbf{\bibinfo{volume}{82}},
  \bibinfo{pages}{2422} (\bibinfo{year}{1999}).

\bibitem[{\citenamefont{Roberts et~al.}(2000)\citenamefont{Roberts, Claussen,
  Cornish, and Wieman}}]{Roberts2000a}
\bibinfo{author}{\bibfnamefont{J.~L.} \bibnamefont{Roberts}},
  \bibinfo{author}{\bibfnamefont{N.~R.} \bibnamefont{Claussen}},
  \bibinfo{author}{\bibfnamefont{S.~L.} \bibnamefont{Cornish}},
  \bibnamefont{and} \bibinfo{author}{\bibfnamefont{C.~E.}
  \bibnamefont{Wieman}}, \bibinfo{journal}{Phys. Rev. Lett.}
  \textbf{\bibinfo{volume}{85}}, \bibinfo{pages}{728} (\bibinfo{year}{2000}).

\bibitem[{\citenamefont{Fedichev et~al.}(1996)\citenamefont{Fedichev, Reynolds,
  and Shlyapnikov}}]{Fedichev1996b}
\bibinfo{author}{\bibfnamefont{P.~O.} \bibnamefont{Fedichev}},
  \bibinfo{author}{\bibfnamefont{M.~W.} \bibnamefont{Reynolds}},
  \bibnamefont{and} \bibinfo{author}{\bibfnamefont{G.~V.}
  \bibnamefont{Shlyapnikov}}, \bibinfo{journal}{Phys. Rev. Lett.}
  \textbf{\bibinfo{volume}{77}}, \bibinfo{pages}{2921} (\bibinfo{year}{1996}).

\bibitem[{\citenamefont{Nielsen and Macek}(1999)}]{Nielsen1999a}
\bibinfo{author}{\bibfnamefont{E.}~\bibnamefont{Nielsen}} \bibnamefont{and}
  \bibinfo{author}{\bibfnamefont{J.~H.} \bibnamefont{Macek}},
  \bibinfo{journal}{Phys. Rev. Lett.} \textbf{\bibinfo{volume}{83}},
  \bibinfo{pages}{1566} (\bibinfo{year}{1999}).

\bibitem[{\citenamefont{Esry et~al.}(1999)\citenamefont{Esry, Greene, and
  Burke}}]{Esry1999b}
\bibinfo{author}{\bibfnamefont{B.~D.} \bibnamefont{Esry}},
  \bibinfo{author}{\bibfnamefont{C.~H.} \bibnamefont{Greene}},
  \bibnamefont{and} \bibinfo{author}{\bibfnamefont{J.~P.} \bibnamefont{Burke},
  \bibfnamefont{Jr.}}, \bibinfo{journal}{Phys. Rev. Lett.}
  \textbf{\bibinfo{volume}{83}}, \bibinfo{pages}{1751} (\bibinfo{year}{1999}).

\bibitem[{\citenamefont{Bedaque et~al.}(2000)\citenamefont{Bedaque, Braaten,
  and Hammer}}]{Bedaque2000a}
\bibinfo{author}{\bibfnamefont{P.~F.} \bibnamefont{Bedaque}},
  \bibinfo{author}{\bibfnamefont{E.}~\bibnamefont{Braaten}}, \bibnamefont{and}
  \bibinfo{author}{\bibfnamefont{H.-W.} \bibnamefont{Hammer}},
  \bibinfo{journal}{Phys. Rev. Lett.} \textbf{\bibinfo{volume}{85}},
  \bibinfo{pages}{908} (\bibinfo{year}{2000}).

\bibitem[{\citenamefont{Leo et~al.}(2000)\citenamefont{Leo, Williams, and
  Julienne}}]{Leo2000}
\bibinfo{author}{\bibfnamefont{P.~J.} \bibnamefont{Leo}},
  \bibinfo{author}{\bibfnamefont{C.~J.} \bibnamefont{Williams}},
  \bibnamefont{and} \bibinfo{author}{\bibfnamefont{P.~S.}
  \bibnamefont{Julienne}}, \bibinfo{journal}{Phys. Rev. Lett.}
  \textbf{\bibinfo{volume}{85}}, \bibinfo{pages}{2721} (\bibinfo{year}{2000}).

\bibitem[{\citenamefont{Chin et~al.}(2000)\citenamefont{Chin, Vuletic, Kerman,
  and Chu}}]{Cheng2000}
\bibinfo{author}{\bibfnamefont{C.}~\bibnamefont{Chin}},
  \bibinfo{author}{\bibfnamefont{V.}~\bibnamefont{Vuletic}},
  \bibinfo{author}{\bibfnamefont{A.~J.} \bibnamefont{Kerman}},
  \bibnamefont{and} \bibinfo{author}{\bibfnamefont{S.}~\bibnamefont{Chu}},
  \bibinfo{journal}{Phys. Rev. Lett.} \textbf{\bibinfo{volume}{85}},
  \bibinfo{pages}{2717} (\bibinfo{year}{2000}).

\bibitem[{Jul()}]{Julienne:2002a}
\bibinfo{note}{Calculations by P. S. Julienne, E. Tiesinga, and C. J. Williams
  (2003), using the model of {\cite{Leo2000}}. Private communication.}

\bibitem[{\citenamefont{Landau and Lifshitz}(1977)}]{LandauLifshitz}
\bibinfo{author}{\bibfnamefont{L.~D.} \bibnamefont{Landau}} \bibnamefont{and}
  \bibinfo{author}{\bibfnamefont{E.~M.} \bibnamefont{Lifshitz}},
  \emph{\bibinfo{title}{Quantum Mechanics: Non-Relativistic Theory}}
  (\bibinfo{publisher}{Pergamon Press}, \bibinfo{address}{Oxford},
  \bibinfo{year}{1977}), \bibinfo{edition}{3rd} ed.

\bibitem[{\citenamefont{Weber et~al.}(2003)\citenamefont{Weber, Herbig, Mark,
  N{\"a}gerl, and Grimm}}]{Weber2003a}
\bibinfo{author}{\bibfnamefont{T.}~\bibnamefont{Weber}},
  \bibinfo{author}{\bibfnamefont{J.}~\bibnamefont{Herbig}},
  \bibinfo{author}{\bibfnamefont{M.}~\bibnamefont{Mark}},
  \bibinfo{author}{\bibfnamefont{H.-C.} \bibnamefont{N{\"a}gerl}},
  \bibnamefont{and} \bibinfo{author}{\bibfnamefont{R.}~\bibnamefont{Grimm}},
  \bibinfo{journal}{Science} \textbf{\bibinfo{volume}{299}},
  \bibinfo{pages}{232} (\bibinfo{year}{2003}), \bibinfo{note}{published online
  5 Dec 2002 (10.1126/science.1079699)}.

\bibitem[{foo({\natexlab{a}})}]{footnote_below10G}
\bibinfo{note}{Below 10~G, the levitation gradient distorts the trapping
  potential too strongly for consistent measurements {\cite{Weber2003a}}.}

\bibitem[{\citenamefont{Chin}(2001)}]{Cheng2001}
\bibinfo{author}{\bibfnamefont{C.}~\bibnamefont{Chin}}, Ph.D. thesis,
  \bibinfo{school}{Stanford University} (\bibinfo{year}{2001}).

\bibitem[{foo({\natexlab{b}})}]{footnote_kineticenergy}
\bibinfo{note}{In the low-temperature limit, the recombination process does not
  depend on the kinetic energy.}

\bibitem[{foo({\natexlab{c}})}]{footnote_iteration}
\bibinfo{note}{Iteration steps are as follows: {I}. Choose values for $\alpha$
  and $\gamma$. {II}. Using these values, calculate a best fit of the solution
  of Eq.~(\ref{diffeqT}) to the temperature measurements, with $T_h$ as single
  fit parameter. {III}. Calculate the error sum of squares $e_N$ between the
  atom number measurements and the solution of Eq.~(\ref{diffeqN}). Steps
  {I}-{III} are repeated while varying $\alpha$ and $\gamma$ to minimize $e_N$
  using a nonlinear optimization algorithm {\cite{Matlab}}.}

\bibitem[{Mat()}]{Matlab}
\bibinfo{note}{{\textsc{matlab}} $6.5$ R13, optimization toolbox, function
  \textit{lsqnonlin}}.

\bibitem[{\citenamefont{Bevington and Robinson}(1992)}]{Bevington}
\bibinfo{author}{\bibfnamefont{P.~R.} \bibnamefont{Bevington}}
  \bibnamefont{and} \bibinfo{author}{\bibfnamefont{D.~K.}
  \bibnamefont{Robinson}}, \emph{\bibinfo{title}{Data reduction and error
  analysis for the physical sciences}} (\bibinfo{publisher}{McGraw-Hill},
  \bibinfo{address}{New York}, \bibinfo{year}{1992}), \bibinfo{edition}{2nd}
  ed.

\bibitem[{foo({\natexlab{d}})}]{footnote_cosc}
\bibinfo{note}{$C$ is predicted to undergo one full oscillation period between
  0 and {$C{_\text{max}}$} when {$a$} varies by a factor of $22.7$
  {\cite{Bedaque2000a}}.}

\bibitem[{foo({\natexlab{e}})}]{footnote_epscorr}
\bibinfo{note}{We include a correction {\cite{Gribakin1993a}} giving
  {$\varepsilon ={\hbar^2}/\left(m(a-\bar{a})^2\right)$}, with
  {$\bar{a}=95.5$~$a_0$} for cesium.}

\bibitem[{\citenamefont{Gribakin and Flambaum}(1993)}]{Gribakin1993a}
\bibinfo{author}{\bibfnamefont{G.~F.} \bibnamefont{Gribakin}} \bibnamefont{and}
  \bibinfo{author}{\bibfnamefont{V.~V.} \bibnamefont{Flambaum}},
  \bibinfo{journal}{Phys. Rev. A} \textbf{\bibinfo{volume}{48}},
  \bibinfo{pages}{546} (\bibinfo{year}{1993}).

\bibitem[{\citenamefont{Takekoshi et~al.}(1998)\citenamefont{Takekoshi,
  Patterson, and Knize}}]{Takekoshi:1998}
\bibinfo{author}{\bibfnamefont{T.}~\bibnamefont{Takekoshi}},
  \bibinfo{author}{\bibfnamefont{B.~M.} \bibnamefont{Patterson}},
  \bibnamefont{and} \bibinfo{author}{\bibfnamefont{R.~J.} \bibnamefont{Knize}},
  \bibinfo{journal}{Phys. Rev. Lett.} \textbf{\bibinfo{volume}{81}},
  \bibinfo{pages}{5105} (\bibinfo{year}{1998}).

\end{thebibliography}

\end{document}